# Predicting Lane Keeping Behavior of Visually Distracted Drivers Using Inverse Suboptimal Control*

Felix Schmitt[1], Hans-Joachim Bieg[1], Dietrich Manstetten[1], Michael Herman[1] and Rainer Stiefelhagen[2]

*Abstract*— Driver distraction strongly contributes to crash-risk. Therefore, assistance systems that warn the driver if her distraction poses a hazard to road safety, promise a great safety benefit. Current approaches either seek to detect critical situations using environmental sensors or estimate a driver's attention state solely from her behavior. However, this neglects that driving situation, driver deficiencies and compensation strategies altogether determine the risk of an accident. This work proposes to use inverse suboptimal control to predict these aspects in visually distracted lane keeping. In contrast to other approaches, this allows a situation-dependent assessment of the risk posed by distraction. Real traffic data of seven drivers are used for evaluation of the predictive power of our approach. For comparison, a baseline was built using established behavior models. In the evaluation our method achieves a consistently lower prediction error over speed and track-topology variations. Additionally, our approach generalizes better to driving speeds unseen in training phase.

## I. INTRODUCTION

### A. Motivation

Driver distraction is a psychological concept that can be defined as "diversification of attention away from activities critical for safe driving towards a competing activity" [1]. Common examples are looking at an in-vehicle display instead of the road ahead of the vehicle (visual distraction), dialing a telephone number which requires removing one hand from the steering-wheel (manual distraction) or conversation (cognitive distraction). According to the U.S. National Highway Traffic Safety Administration 25% of police-reported crashes are caused by inattention in form of distracted or fatigued (e.g. drowsy) drivers [2]. An analysis of the 100-Car Naturalistic Driving Study even concludes that approximately 80% of all crashes and 65% of all near crashes involve inattentive drivers as a contributing factor [3].

Safe driving requires predominately the acquisition and processing of *visual* information for lateral and longitudinal vehicle control [4]. Accordingly, several studies found distinct negative effects of visual distraction, i.e. too long off-road glances, on driving performance, e.g. slower and impaired response to lead vehicle braking [5] or decreased lane-keeping performance [6]. Although conventional driver assistance systems, e.g. forward collision and lane departure warning, can already partially mitigate these decrements, optimal assistance has to consider the driver's distraction state. Ultimately, the driver needs only support if her abilities do not suffice to resolve a hazard. Hence, the knowledge of whether or whether not the driver has currently perceptual insufficiencies can spare her of false alerts or interventions by the assistance system, as demonstrated in [7].

### B. Related work

Motivated by its importance, several algorithms for detection and assessment of inattention have been proposed (reviewed by [8]). Due to progress in real-time estimation of head orientation [9] and eye tracking [10], current approaches that address visual distraction often process the driver's gaze-direction (or the head-orientation as an approximation) for prediction. In this work we, similar as other authors ([3], reviewed by [11]), build on the binary variable $\hat{x}_t^o$ that indicates whether the driver's gaze intersects a so-called Region Of Interest (ROI) ($\hat{x}_t^o = 1$ if and $\hat{x}_t^o = 0$ if not). This region is defined as a plane, through which the driver has to look to gather necessary visual information for lane keeping and headway control in the forward road scenery (Fig. 1).

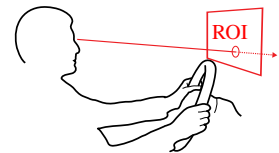

Fig. 1. Intersection of driver's gaze with a potential ROI

An open question in driver distraction research is how to algorithmically decide that the driver is too distracted based on his behavior in a specific situation. In *visual* distraction assessment difficulties arise from the fact that safe driving sometimes even requires to look off-road, e.g. for checking the speedometer. However, an assistance system would warn or even intervene on the algorithms decision. Consequently, if the driver was falsely detected as being critically distracted too frequently the system would become a nuisance and cause the driver to turn it off. Nevertheless, any dangerous situation has to be classified correctly.

Several authors, e.g. [12] and [13], proposed algorithms for decision on critical distraction that in fact detect the periods the driver is engaging into a *secondary* task during the *primary* task of safe driving. In most cases such an additional task of subjective utility and requiring visual attention is the

*This work is part of the public project UR:BAN which is co-funded by the Federal Ministry for Economic Affairs and Energy on basis of a decision by the German Bundestag.

[1]F. Schmitt, H.-J. Bieg, D. Manstetten and M. Herman are with Robert Bosch GmbH, Corporate Research, 70465 Stuttgart, Germany

[2] R. Stiefelhagen is with the Institute for Anthropomatics and Robotics, Karlsruhe Institute of Technology, Germany.

accepted for IEEE Intelligent Vehicles Symposium (IV 2016)

reason for long off-road glances. However, this approach has an issue: Particular visually distracting activities, e.g. selecting a radio station, are commonly socially accepted. These familiar tasks were for example performed in up to 25% of the driving time in one naturalistic driving study [14]. In many situations, they pose little risk, as drivers have developed compensatory mechanisms [15]. If *those* engagements indeed have utility to the driver, then she should rather not receive interventions from the system.

As an alternative to detection of secondary task engagement the subjectively perceived distraction can be predicted [16]. To achieve this, a-posterior distraction rating based on 10s of video of the driver and the outside scenery was used. While this method captures the subjective risk of distraction, it does not reliably relate to objective risk (e.g. probability a crash) which can rapidly change over seconds. Additionally, human ratings are often ambiguous.

Prediction of the *objective* relative crash risk by different ROI-based algorithms was investigated by [11] using data of a naturalistic driving study [3]. Here, similar to the other approaches, the dependence of "safe" gaze behavior on the specific situation is neglected. Furthermore, it is hardly tractable to conduct a study with prototype vehicles containing a similar amount of crash-data for evaluation.

We follow a different idea: Instead of estimating a driver distraction state, we predict both the future in-ROI-state $\hat{x}^o_{t'}$ of the driver and the trajectory of the vehicle as driven by the potentially distracted driver. This allows to assess the current state $\hat{x}^o_t$ wrt. lane keeping by estimating the probability of critical situations caused by driver impairment.

From a methodological point of view, similar inverse optimal control concepts have been applied to other problems of assisted [17] and autonomous driving [18]. However, none of these is suitable for the stochastic and partially observable control problem considered in this work.

### C. Contributions

The main contribution of this work is the proposal of an inverse suboptimal control approach to predict behavior in distracted lane keeping. We consider this as a step towards situation specific visual distraction assessment. Therefore, we introduce the class of partially observable Markov decision processes in II-A and show as a first result in II-B how distracted lane keeping can be modeled as a strategy therein. Next, we explain how the driver's objective inferred by inverse optimal control can be used to predict her behavior in new situations II-C. Maximum causal entropy inverse optimal control [19] is presented in II-D as an approach to account for suboptimal driver behavior. Following, we report the setting of the driving study, we conducted for evaluation in III. As a second contribution we finally present in IV a numerical comparison in terms of predictive power of our approach against a baseline built from Salvucci's well-known two-point steering model [20] and Johnson's model for gaze-allocation [21]. Thereby, we investigate both the overall prediction error and the performance when applying the models to speeds unseen in the training phase.

## II. PROBLEM FORMULATION

### A. Partially Observable Markov Decision Processes

Most optimal control problems belong to the class of Markov Decision Processes (MDPs). An MDP consists of a state $x$, a control $u$, a stochastic process model $\mathcal{P}(x_{t+1}|x_t, u_t)$ and a reward-function $r(x, u)$. The objective of an MDP is to find a stochastic policy/controller $\pi(u|x)$ that maximizes the expected reward $\mathbb{E}[\sum_{t=0}^{T} r(x_t, u_t)|\pi, \mathcal{P}, p_0]$ over a time horizon $T$, conditioned on the policy $\pi$, the dynamics $\mathcal{P}$ and an initial-state distribution $p_0$.

The problem addressed in this work belong to the class of partially observable MDPs (POMDP)s [22], that extend MDPs to states that cannot directly be observed. Here, first the a-posterior distribution $p(x_t|o_{1,...,t})$ of $x_t$ has to be inferred from observations $o_{1,...,t}$ distributed according to an Observation Model (OM) $p_o(o_t|x_t)$. If the a-posterior distribution, the so-called belief $b_t$, is known, the POMDP can be transformed to an equivalent MDP in the belief-state $x' := b$ (see [22] for details).

We motivate the application of POMDPs by the idea that distracted driving is an optimal policy for a *combination* of driving performance *and* utility of an arbitrary secondary task. As both tasks are in conflict, the primary task *by itself* is no longer solved optimally and critical situations can occur. This is the case, especially, if the driver's behavior is only suboptimal as considered in II-D.

### B. Model for Distracted Lane Keeping

Distracted lane keeping can be modeled as an optimal policy $\pi$ in a POMDP in system- and OM-states $(x^s, x^o)^\top$, observations $o$, system- and OM-controls $(u^s, u^o)^\top$ and parametric reward $r = \theta_s^\top \varphi_s(x^s, u^s) + \theta_o^\top \varphi_o(x^o, u^o)$ consisting of:

1) A time-varying linear-affine model of the dynamics of the vehicle in $x^s, u^s$ (system-state dynamics)

$$x^s_{t+1} = A_t x^s_t + B_t u^s_t + a_t + \epsilon_s, \qquad (1)$$

with normal-distributed disturbance $\mathcal{N}(\epsilon_s|\mathbf{0}, \Sigma_s)$. Specifically, we used a discretization of the linear kinematic model given by the ordinary differential equations

$$\underbrace{\begin{bmatrix} \dot{y} \\ \dot{\phi} \\ \dot{\alpha} \end{bmatrix}}_{\dot{x}^s_t} = \underbrace{\begin{bmatrix} 0 & v_t & 0 \\ 0 & 0 & c\, v_t \\ 0 & 0 & 0 \end{bmatrix}}_{\hat{A}_t} \underbrace{\begin{bmatrix} y \\ \phi \\ \alpha \end{bmatrix}}_{x^s_t} + \underbrace{\begin{bmatrix} 0 \\ 0 \\ 1 \end{bmatrix}}_{\hat{B}_t} \underbrace{\dot{\alpha}}_{u^s_t} + \underbrace{\begin{bmatrix} 0 \\ -v_t \kappa_t \\ 0 \end{bmatrix}}_{a_t},$$

with the variables explained in Table I and Fig. 3.

2) A reward $\theta_s^\top \varphi_s$ on $x^s_t$ and $u^s_t$, that encodes the driving objectives of lane keeping. As a driver seeks to keep the vehicle in the center of the road by smooth and minimal steering, the squares of the lane position, the steering angle and the steering velocity are penalized. Also $\dot{y}^2$ is considered, as the lateral velocity relates to the impact severity in case of a crash with an object

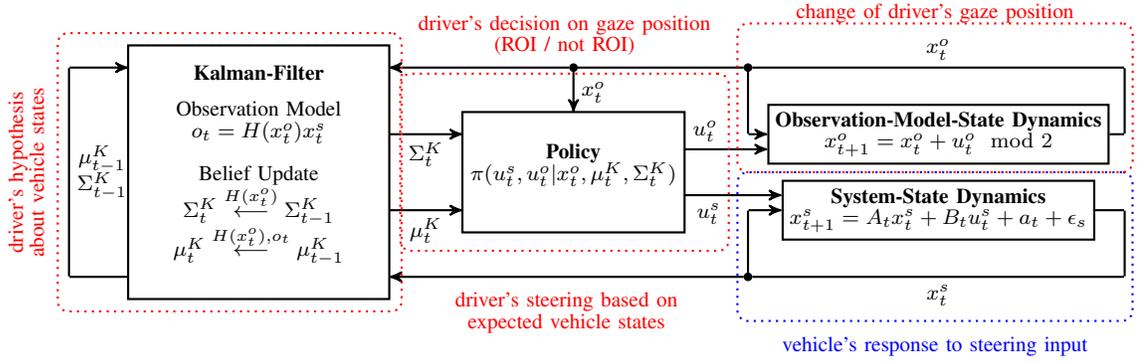

Fig. 2. Control-flow in belief-MDP II-B. Elements of the driver model are indicated in red, while the vehicle model is denoted in blue

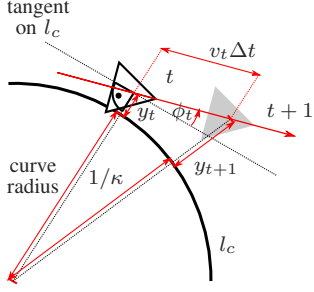

Fig. 3. Change of the vehicle's (denoted by △) position $y_t \to y_{t+1}$ for steering wheel angle $\alpha_t = 0$ and $\phi_t, v_t, \Delta t, \kappa$

TABLE I
VARIABLES OF KINEMATIC VEHICLE MODEL

|   | Definition | Unit |
|---|---|---|
| $y$ | lateral position wrt. lane center $l_c$ | m |
| $\phi$ | angle between tangent of lane $l$ and vehicle's longitudinal axis | rad |
| $\alpha$ | steering-angle | rad |
| $v$ | vehicle's absolute velocity | m/s |
| $\kappa$ | curvature of lane | 1/m |
| $c$ | steering-wheel transmission ratio | |

on the neighboring lane. Altogether, this results in a reward model

$$\theta_s^\top \varphi_s(x_t^s, u_t^s) = -\theta_1 y_t^2 - \theta_2 \dot{y}_t^2 - \theta_3 \alpha_t^2 - \theta_4 \dot{\alpha}_t^2. \quad (2)$$

3) A model of the driver's impairment during visual distraction. Thereby, the OM-state $x_t^o$ is nothing but the in-ROI indicator $\hat{x}_t^o$ from I-B. For the purpose of this work the attentive driver is assumed to perfectly perceive all vehicle states, while the driver engaged in the secondary task is assumed to not perceive $y, \phi$ but sense $\alpha$ with one hand on the steering wheel (see Fig. 4). This can be mathematically implemented by the $x_t^o$-dependent observation model

$$o_t = H(x_t^o)x_t^s \;, \; H(x_t^o) = \begin{cases} e_3 e_3^\top & x_t^o = 0 \\ I & x_t^o = 1 \end{cases} \quad (3)$$

where $I$ is the identity matrix and $e_3 = (0,0,1)^\top$. The driver's ability to switch his gaze between the secondary task and the road at will, is incorporated by the OM-control $u_t^o \in \{0,1\}$ and the OM-state dynamics

$$x_{t+1}^o = (x_t^o + u_t^o) \mod 2. \quad (4)$$

4) A reward $\theta_o^\top \varphi_o(x_t^o, u_t^o)$ to account for the secondary task utility and costs for switching the gaze. We assume a constant utility for every time-step of engagement by progress in the task completion and a constant penalty for gaze-switches. This leads to the reward function

$$\theta_o^\top \varphi_o(x_t^o, u_t^o) = \theta_5 \mathbb{I}_{x_t^o=1} + \theta_6 \mathbb{I}_{u_t^o=1} \quad (5)$$

where $\theta_5 \geq 0$, $\theta_6 \leq 0$ and $\mathbb{I}_{(.)}$ is the indicator ($\mathbb{I}_\chi = 1$ for a true expression $\chi$, $\mathbb{I}_\chi = 0$ otherwise).

Because of the linearity in (1) and (3) this POMDP can easily be transformed into its belief-MDP: The belief $b_t$ in $x_t^s$ is fully specified by the a-posterior expectation $\mu_t^K$ and the a-posterior covariance $\Sigma_t^K$, resulting from the corresponding Kalman-filter. Fig. 2 visualizes the final belief-MDP.

### C. Inverse Optimal Control for Behavior Prediction

To predict future system states, the presented POMDP process model has to be combined with a model of the driver's policy $\pi$. In this work we indirectly infer $\pi$ from data $\mathcal{D} = \{(x, u)_i\}$, by estimating the parameters $\theta$ of the driver's reward $r(x, u) = \theta^\top \varphi(x^s, x^o, u^s, u^o)$ under the assumption of optimal behavior. This approach is referred to as Inverse Optimal Control (IOC). Once the reward is known, the driver's behavior can be predicted on arbitrary $v_t, \kappa_t$ by computation of the policy based on an optimality criterion. Compared to direct estimation of $\pi$, IOC has the advantage that the parameters $\theta$ are less dependent on the external model variables $v_t, \kappa_t$, as they model driver preferences, potentially applying globally. We empirically investigate this hypothesis in Sec. IV.

### D. Maximum Causal Entropy Inverse Optimal Control

In our case of a linear combination of reward features $r(x, u) = \theta^\top \varphi(x, u)$, matching the empirical feature expectation

$$\mathbb{E}[\varphi(x,u)|\mathcal{D}] = \mathbb{E}\left[\sum_{t=0}^T \varphi(x_t, u_t)\Big|\pi_\dagger, \mathcal{P}, p_0\right] \quad (6)$$

by the optimal policy $\pi_\dagger$ of reward $\theta_\dagger^\top \varphi(x,u)$ results in:

$$\mathbb{E}[\theta_\dagger^\top \varphi(x,u)|\mathcal{D}] = \mathbb{E}\Big[\sum_{t=0}^{T} \theta_\dagger^\top \varphi(x_t, u_t)\Big|\pi_\dagger, \mathcal{P}, p_0\Big], \quad (7)$$

hence $\theta_\dagger^\top \varphi(x,u)$ is a reward for which the observed behavior is optimal [23]. Although many approaches to IOC in MDPs have been proposed [24] not all are suitable for modeling real-world driving behavior. The reason is that humans often deviate from the optimal behavior [25] what needs to be taken into account in modeling distracted driving. Therefore, we apply the Maximum Causal Entropy (MCE) method [19]. Here, a stochastic policy $\pi(u|x)$ of maximum causal entropy, that fullfills condition (6), is computed by the following optimization problem:

$$\max_{\pi(u|x)} -\mathbb{E}\Big[\sum_{t=0}^{T} \log\big(\pi(u_t|x_t)\big)\Big|\pi, \mathcal{P}, p_0\Big] \quad (8)$$

$$\text{s.t.} \quad \mathbb{E}[\varphi(x,u)|\mathcal{D}] = \mathbb{E}\Big[\sum_{t=0}^{T} \varphi(x_t, u_t)\Big|\pi, \mathcal{P}, p_0\Big]. \quad (9)$$

The rationale behind the maximization of the casual entropy of $\pi$ is that the resulting policy effectively deviates least from the true behavior $\pi^\star$ if a worst-case $\pi^\star$ is assumed (details in [19]). Hence, the estimation is very robust.

*1) Suboptimal policies from MCE:* Although reward parameters $\theta$ are not explicitly estimated in MCE, it is considered an inverse optimal control approach due to the following property: The likelihood of the estimated $\pi$ to act like any deterministic policy $u = \bar\pi(x)$ is a monotonic function of

$$\mathbb{E}\Big[\sum_{t=0}^{T} \hat\theta^\top \varphi(x_t, u_t)\Big|\bar\pi, \mathcal{P}, p_0\Big], \quad (10)$$

where $\hat\theta$ is the Lagrangian multiplier of constraint (9) [26](c.6). That means the higher the expected value of $\hat\theta^\top \varphi(x,u)$ the likelier is the behavior according to $\bar\pi$. Consequently, $\pi$ is interpreted as a suboptimal, yet goal-oriented policy, with respect to the reward model $\hat\theta^\top \varphi(x,u)$.

*2) Parameter learning in MCE-IOC:* The optimization problem (8)-(9) is solved for optimal $\pi_\star$ by gradient descent on its Lagrangian-dual problem in $\theta$ [19]:

1) Calculate $\pi$ for current $\theta$ using the recursion

$$\tilde Q(u_T, x_T) = \theta^\top r(u_T, x_T), \quad (11)$$
$$V(x_{t+1}) = \log \int \exp\big(\tilde Q(u_{t+1}, x_{t+1})\big)\, du_{t+1},$$
$$\tilde Q(u_t, x_t) = \theta^\top r(u_t, x_t) + \mathbb{E}\big[V(x_{t+1})|\mathcal{P}\big],$$
$$\pi(u_t|x_t) = \exp(\tilde Q(u_t, x_t)) / \int \exp(\tilde Q(\tilde u_t, x_t))\, d\tilde u_t.$$

2) Simulate $\pi$ in the MDP starting from $p_0$ to compute

$$E = \mathbb{E}\Big[\sum_{t=0}^{T} \varphi(x_t, u_t)\Big|\pi, \mathcal{P}, p_0\Big] \quad (12)$$

3) Update $\theta \leftarrow \theta - \eta\big(\mathbb{E}[\varphi(x,u)|\mathcal{D}] - E\big)$ using a suitable step-size $\eta$.

## III. COLLECTED DATA IN DRIVING STUDY

### A. Experiment

To evaluate our modeling approach, we conducted a study on a public highway. We decided against a study in a driving simulator because of possible influences on the participants' behavior by absence of real risk. In addition to that, evaluation of prediction robustness on realistic sensor-input, i.e. noisy signals, is important.

We recruited seven drivers (6 male, 1 female). All drivers took part in a driving safety training prior to the experiment. The experiment consisted of four fixed driving speed conditions $\{80, 90, 100, 110\}$ km/h. Vehicle speed was controlled by the vehicle's Adaptive Cruise Control (ACC) to prevent drivers from adjusting their speed as a compensatory action while being engaged in a secondary task. A conservative time gap was employed to ensure that the distance to preceding vehicles did have the least possible influence on the drivers' behavior. When the vehicle traveled at the required speed, the measurement periods were started. Such a period was either a reference, where the driver had to keep the lane fully attentive or involved a visually distracting secondary task. At each speed three secondary tasks and three reference periods per participant were triggered by the investigator.

The task we used in this experiment consisted of incrementally reading a series of 30 random numbers $n_i \in \{1, 2\}$ from a display and typing them in a number pad (see Fig. 4). Thereby, the participants were instructed to *"perform the*

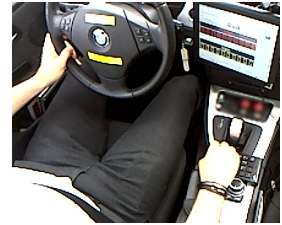

Fig. 4. Artificial secondary task used in the experiment

*secondary task as quickly and correctly as possible while not endangering the driving safety"*. This artificial task was chosen, as it resembles the principle of a variety of real visual-manual tasks performed while driving, and possesses several advantages. First, the task state is fully measurable and can be modeled easily, in contrast to the vehicle's interaction system. Second, the participants needed only little practice to reach maximum execution performance, resulting in no significant learning effects during the experiment.

### B. Recorded Data

We used a series mono-camera system for tracking the lane boundaries and recorded the position of the lane, the angle between tangent of the lane boundary and the vehicle's longitudinal axis and the curvature via the vehicles Controller Area Network (CAN). A commercial three-camera infra red eye tracking system with active illumination was used for estimation of the driver's gaze direction, while an own ROI-based algorithm detected whether the driver's gaze was off or on the road. Steering wheel position and velocity as well as absolute velocity were also recorded via the CAN. Hence, beside the eye-tracking systems we relied solely on signals that are already accessible in today's series-production vehicles.

*C. Preprocessing*

In order to ensure a quality dataset for the numerical evaluation, the following preprocessing and filtering steps were performed on the collected raw data:

*1) Selection of valid trials according to protocol:* We automatically excluded lane changes and their preparation phases. As a different driving maneuver than lane keeping it requires a different driving and gaze policy. Also situations where the ACC controller reduced the vehicle speed by more than 5% were left out due to possible influence on the drivers' behavior. The final dataset consisted of 136 valid segments comprising reference and secondary task periods with an average duration of $\approx 50\,\text{s}$.

*2) Sub-sampling and filtering of signals:* We first sub-sampled the variables of the vehicle model to 25 Hz. Thereafter, the Rauch-Tung-Striebel-smoother [27] was employed for filtering of the signals using the kinematic vehicle model (1). The system parameters $c, \Sigma_s$ as well as the sensor noise-covariance $\Sigma_m$ were estimated by an expectation-maximization approach as proposed for vehicle models by [28].

## IV. EVALUATION

To demonstrate the effectiveness of our IOC-based approach we conducted two numerical evaluations. Here we compared against a baseline comprising of established behavior models.

*A. Baseline*

[21] presented a model for gaze-allocation in visual dual-tasking, where the probability of a glance to a task is a logistic function of the uncertainty in its states. In our case uncertainty is only present in the vehicle states - the random number $n_i$ is either known to the driver if she has seen it on the display or unknown else. Therefore, we applied the following variant of the original approach

$$p(u_t^o | x_t^o = 1) = \frac{\mathbb{I}_{u_t^o=0}\exp(\lambda_1) + \mathbb{I}_{u_t^o=1}}{\exp(\lambda_1) + 1} \quad (13)$$

$$p(u_t^o | x_t^o = 0) = \frac{\mathbb{I}_{u_t^o=0}\exp(\lambda_2 + \text{tr}(\Lambda \Sigma_t^K)) + \mathbb{I}_{u_t^o=1}}{\exp(\lambda_2 + \text{tr}(\Lambda \Sigma_t^K)) + 1}. \quad (14)$$

Here $\lambda_1, \lambda_2, \Lambda$ are parameters and $\text{tr}(.)$ is the sum of the matrix diagonal elements.

For modeling human predictive steering we adopted the *two-point-steering model* of [20]. In this model the driver steers by means of a visual near-angle $\beta_1$ and a visual far-angle $\beta_2$. In a lane $l$ of width $w$, $\beta_1$ is defined as the angle between the straight through the vehicle's center and a near point on the lane center $l_c$ and the vehicle's longitudinal axis. $\beta_2$ is defined as the angle of minimal magnitude between the tangents from the vehicle's center to both boundaries $l_b$ and the vehicle's longitudinal axis (see Fig. 5).
We computed $\beta_1, \beta_2$ using the approximations

$$\beta_1 = -\arctan\left(y\,(2\,\text{m})^{-1}\right) - \phi \quad (15)$$

$$\beta_2 = \begin{cases} -\arccos\left((1 - \kappa\,(w/2+y))^{-1}\right) - \phi & \kappa < 0 \\ +\arccos\left((1 + \kappa\,(w/2-y))^{-1}\right) - \phi & \kappa \geq 0 \end{cases} \quad (16)$$

and fed both angles and the steering-wheel angle into the stochastic policy $\dot{\alpha} = k_1\beta_1 + k_2\beta_2 + k_3\alpha + \epsilon_k$ with parameters $k_1, k_2, k_3$ and a normally-distributed random-variable $\epsilon_k$ ([20] used a PD-controller that was, however, less stable in combination with our approximation due to jumps in $\beta_2$). Both models were linked by replacing $y, \phi$ in (15),(16) with their expectations if $x_t^o = 0$, i.e. when the modeled driver looks off the road, similar to [29].

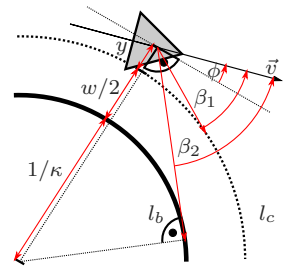

Fig. 5. Illustration of the variables of the two-point steering model.

*B. Evaluation protocol*

For the numerical evaluation we further subdivided all valid segments into snippets of $\approx 5\,\text{s}$ (overlap $\approx 2.5\,\text{s}$) to account for a realistic prediction horizon in a real-time system.
In the training-phase of the MCE-IOC approach we relaxed the feature matching condition (6) to

$$\frac{\left|\mathbb{E}[\varphi(x,u)|\mathcal{D}]_i - \mathbb{E}\left[\sum_{t=0}^T \varphi(x,u)\Big|\pi,\mathcal{P}\right]_i\right|}{\left|\mathbb{E}[\varphi(x,u)|\mathcal{D}]_i\right|} \leq 1\%, \quad (17)$$

(equivalent to a $L1$ regularization of $\theta$) to prevent over-fitting. The parameters of the baseline were inferred using $L1$-regularized maximum likelihood estimation for generalized linear models as implemented in MATLAB's `lassoglm` function [30].
In every interval 100 trajectories $x^s, x^o$ were sampled for each model, starting in the first state of the segment. Thereby, we incrementally simulated the driver's behavior $\pi$ and the responses of the POMDP model (II-B) using the pre-estimated parameters $c, \Sigma_s$ from III-C. On the obtained data the following metrics were computed:

- Expected squared error of lane-position

$$\text{SE}(y) = \mathbb{E}\Big[\frac{1}{T}\sum_{t=0}^T (y_t - y_t^{\mathcal{P}})^2 | \pi, \mathcal{P}, p_0\Big] \quad (18)$$

- Kullback-Leiber divergence between empirical off-road gaze duration $d$ and prediction of the model

$$\mathrm{KL}(p_\mathcal{D}||p) = \sum_d p_\mathcal{D}(d)\Big(\log\big(p_\mathcal{D}(d)\big) - \log\big(p(d)\big)\Big) \quad (19)$$

*1) Overall prediction performance:* We first evaluated the overall prediction performance by splitting the dataset into a training set and a test set of equal size randomly and independently of driver, velocity and track-topology. Afterwards the roles of the datasets were swapped. For better estimation of the error statistics this 2-fold cross-validation procedure was repeated 10 times.

*2) Transfer performance:* To investigate the generalization quality on unseen velocities we conducted a second evaluation. Here we trained on a random selection of half of the data of one single speed condition. We thereafter tested on the remaining half with the same and on all data of other velocities. In this evaluation we performed 5 repetitions of 2-fold cross-validation.

### C. Results

The results of IV-B.1 are summarized in Table II. Due to the skew-shaped error distributions (see Fig.6) we report the total median (instead of the mean) in training and test of both methods in both metrics.

TABLE II
OVERALL PREDICTION PERFORMANCE

| Metric | Baseline | | Ours | |
|---|---|---|---|---|
| | Train | Test | Train | Test |
| SE | 0.0482 | 0.0484 | 0.0154 | 0.0157 |
| KL | 0.0998 | 0.1006 | 0.0731 | 0.0727 |

Table III shows the errors of IV-B.2. We present the total median errors of the 10 evaluations. Thereby, the results of our approach are in **bold** letters.

TABLE III
TRANSFER PERFORMANCE

| Train | | Test | | | |
|---|---|---|---|---|---|
| | | 80 km/h | 90 km/h | 100 km/h | 110 km/h |
| 80 km/h | SE | **0.0154** | 0.0143 | 0.0164 | 0.0154 |
| | | 0.0293 | 0.0572 | 0.0725 | 0.0836 |
| | KL | **0.0851** | 0.0744 | 0.0738 | 0.0701 |
| | | 0.1280 | 0.1180 | 0.1025 | 0.1216 |
| 90 km/h | SE | 0.0152 | **0.0148** | 0.0159 | 0.0163 |
| | | 0.0310 | 0.0365 | 0.0477 | 0.0627 |
| | KL | 0.0833 | **0.0748** | 0.0730 | 0.0712 |
| | | 0.1307 | 0.1146 | 0.1018 | 0.1200 |
| 100 km/h | SE | 0.0147 | 0.0160 | **0.0155** | 0.0161 |
| | | 0.0311 | 0.0390 | 0.0450 | 0.0584 |
| | KL | 0.0838 | 0.0763 | **0.0731** | 0.0693 |
| | | 0.13743 | 0.1124 | 0.1100 | 0.1173 |
| 110 km/h | SE | 0.0157 | 0.0152 | 0.0179 | **0.0177** |
| | | 0.0401 | 0.0408 | 0.0479 | 0.0603 |
| | KL | 0.0825 | 0.0724 | 0.0755 | **0.0721** |
| | | 0.1249 | 0.1076 | 0.0901 | 0.1102 |

Fig. 6 depicts the error distributions of both evaluations.

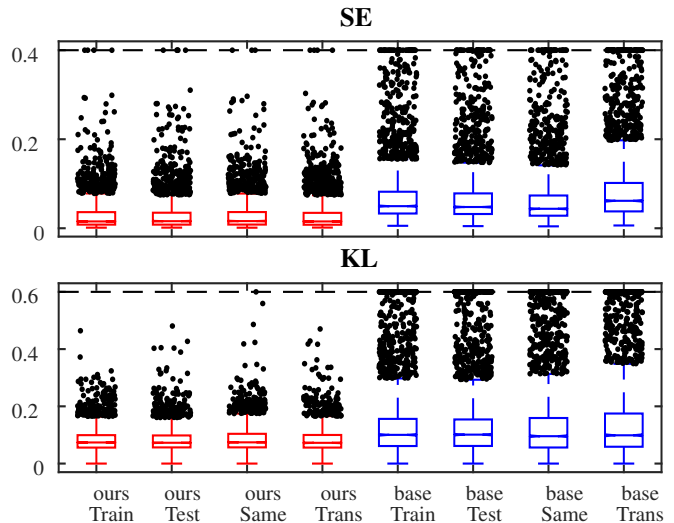

Fig. 6. Box indicates the $[25, 75]$-% interval $\mathcal{I}$ with the median as horizontal line. Data whose distance to the median exceeds $1.5\times$ distance median to corresponding quantile, is considered outliers (·). *Train* and *Test* denote the conditions of IV-B.1, while *Same* denotes the test error on the same speed and *Transfer* the test error on the other unseen speeds of IV-B.2.

### D. Discussion

IV-B.1 shows significantly lower prediction error of our approach based on MCE-IOC compared to the baseline (Wilcoxon signed-rank $p_\text{test} \leq 10^{-12}$). As training and test error for both methods and in both metrics differed only slightly (Wilcoxon ranksum $p_\text{test} \geq 0.01$), adequate regularization was chosen for both methods. A possible explanation for the better SE of our approach can be that the explicit incorporation of the vehicle's velocity and the future track curvature allows more precise trajectory prediction. Although Johnson's barrier model already incorporates the growing uncertainty over time it might have worse KL than the MCE-IOC approach because it lacks a direct link to lane keeping controller.

IV-B.2 gives further insight into the performance differences: MCE-IOC performed significantly better in both metrics ($p_\text{test} \leq 10^{-12}$) over all conditions except for KL Train 110 km/h Test 100 km/h ($p_\text{test} = 0.01$). Although a test on between-condition differences revealed significant variations (Kruskal-Wallis $p_\text{test} \leq 10^{-5}$), these were more pronounced in the baseline ($p_\text{test} \leq 10^{-12}$ vs. $p_\text{test} \geq 10^{-6}$ in MCE-IOC) what could be validated by significantly lower variance (Ansari-Bradley $p_\text{test} \leq 10^{-12}$).

The hypothesis that the driver preferences $\theta$ estimated by MCE-IOC are better transferable than a directly estimated policy could be confirmed *in this experiment* by IV-B.2: The difference between the error on unseen speeds and the median test error on seen speeds was significantly higher for the baseline approach in both metrics (Wilcoxon ranksum SE $p_\text{test} = 10^{-11}$, KL $p_\text{test} = 10^{-9}$). However, the general hypothesis has to be further investigated on significantly lower speeds in future work.

## V. CONCLUSION

We presented a inverse optimal control based method to predict the behavior of potentially distracted driver in lane keeping. Here, the key elements are a model of goal-oriented driver behavior wrt. a reward, an explicit model of the driver's impairment and a dynamics model of the vehicle responses. Once the reward is inferred from observed behavior, it can be transferred to new situations for estimation of likely driver behavior what was empirically validated for unseen driving speeds. In the evaluation we also showed lower prediction error of our method compared to a baseline comprising of domain-specific models [21],[20].

Our approach can be used for situation-dependent assessment of visual driver distraction wrt. lane keeping, by prediction of the probability of critical incidents, i.e. lane-departure. However, for deployment in a final distraction mitigation system it has to be completed with prediction of lateral driving behavior. Hence, in future work we will address inverse optimal control approaches for modeling speed adjustment and headway control of distracted drivers. Additionally, we also want to incorporate estimation of observation models II-B, eq. (3) into the inference procedure. This is necessary for behavior prediction in secondary task, e.g. reading the speedometer, where the driver can to some extent perceive the road by peripheral vision. Here, the knowledge of sub-optimal behavior of the driver can also be exploited [31].